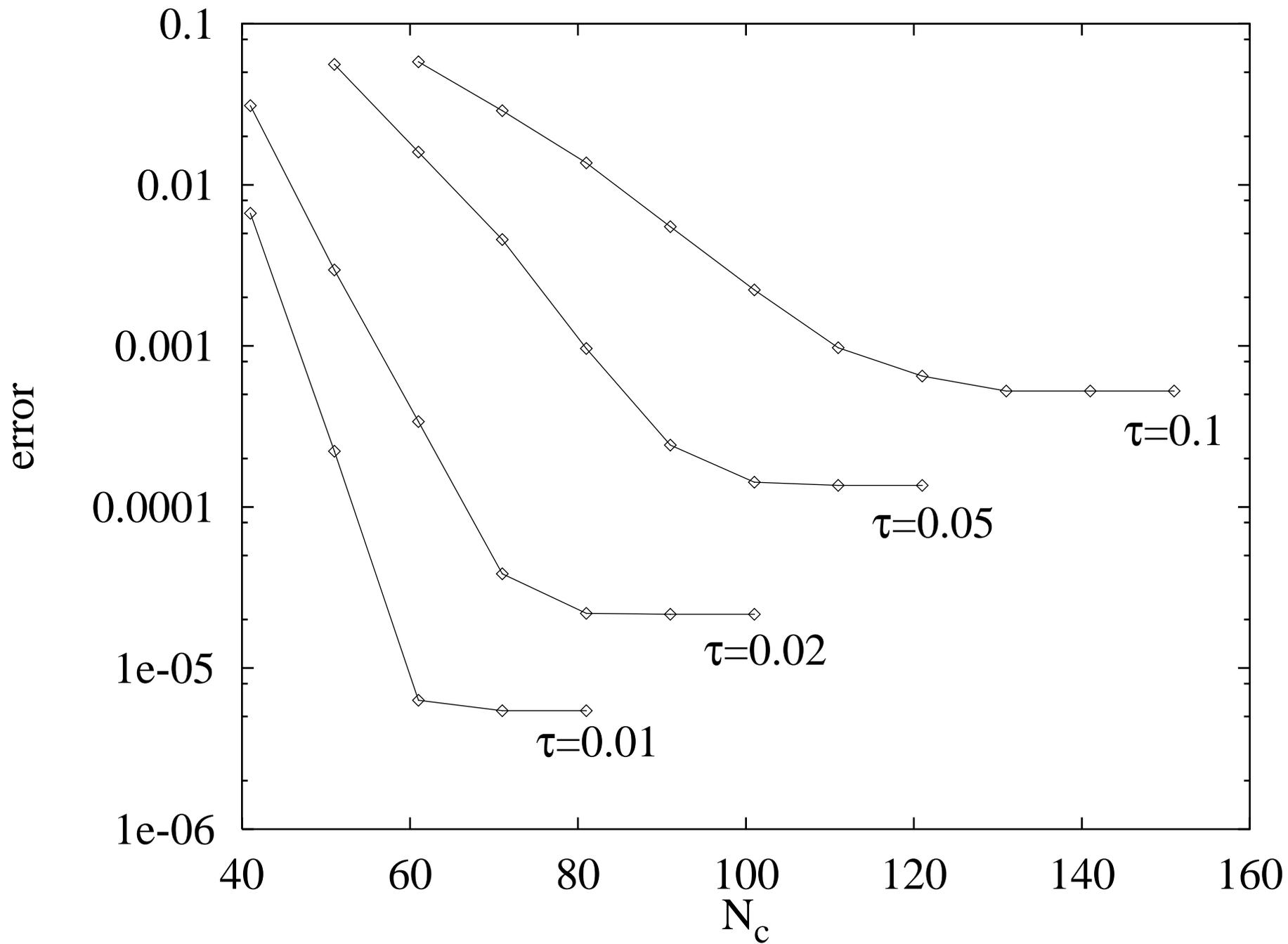

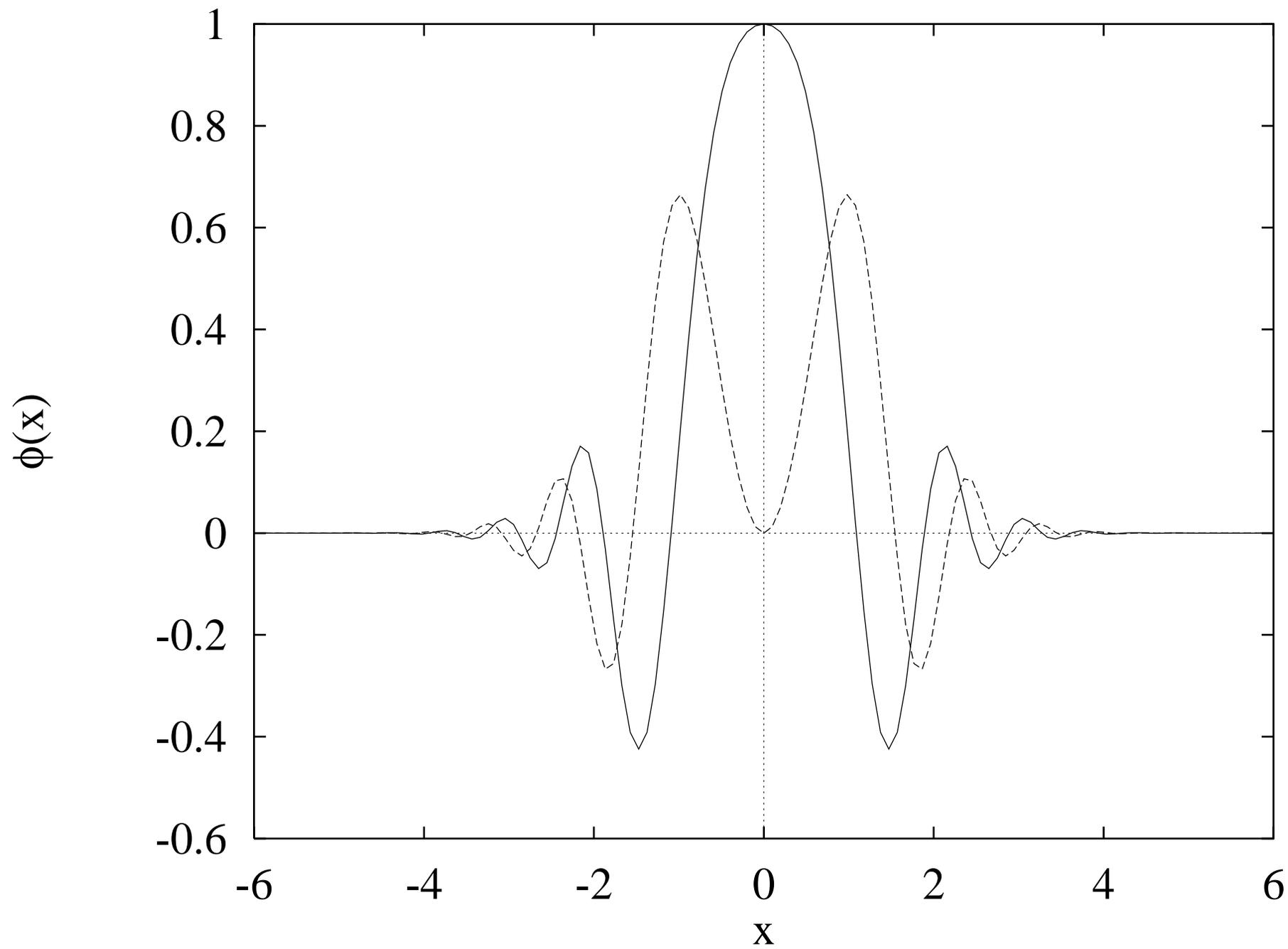

# A Novel Integration Scheme for Partial Differential Equations: an Application to the Complex Ginzburg-Landau Equation

Alessandro Torcini, Helge Frauenkron and Peter Grassberger
*Theoretische Physik, Bergische Universität-Gesamthochschule Wuppertal,*
*D-42097 Wuppertal, Germany*
(November 9, 1995)

## Abstract

A new integration scheme, combining the stability and the precision of usual pseudo-spectral codes with the locality of finite differences methods, is introduced. It turns out to be particularly suitable for the study of front and disturbance propagation in extended systems. An application to the complex Ginzburg-Landau equation shows the higher precision of this method with respect to spectral ones.

# I. INTRODUCTION

The integration of partial differential equations (PDE's) is a relevant topic by itself, but efficient and accurate algorithms are particularly important for extended systems exhibiting spatio-temporal chaos. Usually, PDE's are integrated according to one of two general schemes: finite difference methods or time splitting pseudo-spectral codes [1]. In the present contribution we will introduce a new algorithm that will combine the accuracy and stability of the spectral method with the locality property of finite differences schemes.

The present method can be applied to any PDE, be it Hamiltonian or dissipative. For most problems, it should be comparable in efficiency to time splitting pseudo-spectral codes, but there is one class of problems where it seems superior to any previously known method.

This is the propagation of disturbances or fronts into *unstable* states. The classic example is the propagation of fronts in the KPP (or Fisher-) equation [2]. Similar are propagation of solidification fronts in directional solidification and viscous fingering [3,4]. Another phenomenon which is closely related mathematically though it seems at first rather different, is propagation of disturbances in spatially extended chaotic systems [5,6]. In these cases, spectral methods cannot be used because of the instability of the state into which the front or the disturbance propagates. Due to the non-locality of the Fourier transform, it is in general impossible to keep the state completely unperturbed in front of the perturbation. Instead, there will be small perturbations in regions which should not have been reached yet by the front (at least due to round-off errors), and these perturbations will grow exponentially due to the instability of the state, rendering any precise measurement of the front properties impossible.

For sake of brevity, we will discuss in this article only an application to propagation of disturbances in the one-dimensional complex Ginzburg-Landau equation (CGLE). We will consider the 1-d CGLE because it represents a paradigmatic example for a large class of spatially extended systems [3] exhibiting, for different parameter intervals, behavior that ranges from spatio-temporal chaos [7] to stable and intermittent regimes [8]. Moreover, also the nonlinear Schrödinger equation can be recovered as a particular limit of the CGLE. Propagation turns out to play a crucial rôle in the understanding of the dynamics of spatially extended chaotic systems [4–6,9]. It is particularly important in systems with so-called "chaotic supertransients" [10], where it is responsible for the chaos observed in infinite systems.

The paper is organized as follows. In the next section the integration scheme will be described in detail together with a specific implementation of the algorithm for the CGLE. Some accuracy tests are also reported in Sec. II. Section III is devoted to a comparison between propagation speeds measured with our algorithm, and speeds obtained from the Lyapunov analysis of [6]. Finally, Sec. IV summarizes the main results of the work, along with a few concluding remarks.

# II. THE INTEGRATION ALGORITHM

In the present article we will limit ourselves to PDE's of the form

$$\frac{\partial A(x,t)}{\partial t} = \mathcal{L} A(x,t) = (\mathcal{T} + \mathcal{V}) \ A(x,t) \tag{1}$$

where $A(x,t)$ is a 1 component field, and $x \in [0,L]$ is 1-dimensional. We assume that the operator $\mathcal{L}$ can be split into two parts, a linear operator $\mathcal{T}$ containing all the spatial derivatives and a strictly local operator $\mathcal{V}$ which is in general non-linear. Both $\mathcal{T}$ and $\mathcal{V}$ are assumed to be translation invariant. Generalizations to multi-component fields and to higher dimensions are straightforward.

Formally the solution of Eq. (1) can be expressed as

$$A(x,t) = e^{t\mathcal{L}} A(x,0) . \tag{2}$$

We assume that $\mathcal{T}$ and $\mathcal{V}$ do not commute with each other. Thus, one cannot factorize the exponential, but should use the Baker-Campbell-Hausdorff (BCH) formula to express $e^{t(\mathcal{T}+\mathcal{V})}$ as a product of terms $e^{td_i\mathcal{T}}$ and $e^{tc_i\mathcal{V}}$. An appropriate choice of the coefficients $d_i$ and $c_i$ can reduce noticeably the errors committed in neglecting higher order commutators in the BCH expression [11–13]. For sake of simplicity, we will consider the lowest order approximation [14] corresponding to the Trotter formula

$$e^{t\mathcal{L}} = e^{t\mathcal{V}/2} \cdot e^{t\mathcal{T}} \cdot e^{t\mathcal{V}/2} + \mathcal{O}(t^3) . \tag{3}$$

This belongs to the class of the so-called time splitting algorithms [1] and it is commonly referred as "leap-frog method". In practice, when performing this repeatedly, it is advantageous to combine the two half steps in all but the first and last iterations. Except for the latter, each iteration bringing us from $t = n\tau$ to $t' = (n+1)\tau$ corresponds then to two successive integration steps,

$$A_{n+1/2}(x) = e^{\tau \mathcal{V}} A_n(x) \quad ; \quad A_{n+1}(x) = e^{\tau \mathcal{T}} A_{n+1/2}(x) \quad ; \tag{4}$$

The first step, containing no spatial derivatives, can easily be done in coordinate space where it consists just of a multiplication or of the solution of a simple ODE. The second step would be non-local in coordinate space, and is thus usually done in momentum space. Thus, the field $A$ is first Fourier transformed so that the integration takes the form of a simple multiplication,

$$\tilde{A}_{n+1}(p) = e^{\tau \tilde{T}(p)} \tilde{A}_n(p) , \tag{5}$$

and an inverse Fourier transform brings the field back into coordinate representation.

The novelty of the our method, compared to the usual time splitting pseudo-spectral scheme described above, consists just in performing the second step also in configuration space, avoiding thereby the two Fourier transforms. The rest of the integration algorithm is left unchanged. If we neglect spatial discretization, we would have

$$A_{n+1}(x) = \int_{-\infty}^{\infty} \phi(\xi) A_{n+1/2}(x - \xi) \tag{6}$$

with

$$\phi(\xi) = \frac{1}{2\pi} \int dp \ e^{ip\xi + \tau \tilde{T}(p)} \ . \tag{7}$$

The crucial observation is that $\phi(\xi)$ will be strongly centered at $\xi \approx 0$ if the step $\tau$ is small. Thus, when using a spatial grid, the integration can be replaced by a sum which involves only very few terms. Let us assume that we can neglect $\phi(\xi)$ outside an interval $[-S, S]$, and that we work with a grid size $\Delta x$. Then each application of eq.(6) involves $N_c = (2S+1) \cdot \Delta x$ operations (the values of $\phi(i\Delta x)$ are of course computed only once during the initialization of the routine). This is to be compared to the $\mathcal{O}(2 \ln[L/\Delta x])$ operations for the two FFT's in the spectral method.

To be specific, let us now consider the CGLE

$$A_t = (1 + ic_1)A_{xx} + A - (1 - ic_3)|A|^2 A \tag{8}$$

where the field $A(x,t)$ is complex, whereas the parameters $c_1$ and $c_3$ assume real positive values. The analysis reported in this paper will be limited to a parameter region where it is known that Eq. (8) will show a chaotic behaviour [7] (namely, $c_1 = 3.5$ and $0.6 \leq c_3 \leq 0.9$).

The integration involving the operator $\mathcal{V}$ can be performed rewriting the field as $A(x,t) = \rho(x,t)e^{i\psi(x,t)}$ and solving the ordinary differential equations

$$\frac{\partial \psi(x,t)}{\partial t} = c_3 \ \rho^2(x,t) \tag{9}$$

and

$$\frac{\partial \rho^2(x,t)}{\partial t} = 2[\rho^2(x,t) - \rho^4(x,t)] \ . \tag{10}$$

The solution of eq.(10) is given by

$$\rho(x,\tau) = [e^{-2\tau}(1/\rho(x,0)^2 - 1) + 1]^{-1/2} \ . \tag{11}$$

Inserting this into eq.(9) gives

$$\psi(x,\tau) = \psi(x,0) + c_3 \left\{ \tau + \ln[\rho(x,0)/\rho(x,\tau)] \right\} \ . \tag{12}$$

Since $\tilde{T}(p)$ is quadratic in $p$, the kernel for the second integration step reads simply

$$\phi(x) = \sqrt{\frac{\beta}{\pi}} e^{-\beta_r x^2} [\cos(\beta_i x^2) - i \sin(\beta_i x^2)] \tag{13}$$

where $\beta = \beta_r - i\beta_i = (1 - ic_1)/[4\tau(1 + c_1^2)]$. In fig.1 the real and complex part of $\phi(x)$ are shown for a typical choice of parameters. It is clear that the shape of $\phi(x)$ will strongly depend on the chosen time step and on the parameter $c_1$ in the CGLE. In order to achieve sufficient accuracy, the shape and the details of $\phi(x)$ should be well resolved within the chosen spatial resolution $dx$. In fig.2 we show average errors accumulated during a fixed total integration time $T = 0.2$ for several values of $\tau$ and of the number $N_c$ of terms in the convolution. In these simulations, the spatial resolution and the parameters of the CGLE were kept fixed at $\Delta x = 0.098$, $c_1 = 3.5$ and $c_3 = 0.9$.

We see in particular that the errors saturate as $N_c$ increases, and we verified that the limits of the errors for large $N_c$ coincide exactly with the errors of the spectral method, as it should be the case. Assume that we want to tolerate a total error which is, say, a factor 2 larger than that of the spectral method. Then we can read off from fig.2 that the necessary values of $N_c$ are $\approx 57$ (for $\tau = 0.01$), 70 (for $\tau = 0.02$), 90 (for $\tau = 0.05$), and 110 (for $\tau = 0.1$). On the other hand, using the FFT routine C06ECF of the NAG library and grid sizes $N$ which are powers of 2, we found that our algorithm (whose speed is essentially proportional to $N_c$) needed more CPU time by a factor $0.15 N_c / \ln N$, for $N > 2048$. Thus it seems that our algorithm with $\tau = 0.01$ and $N_c = 57$ has about the same speed and precision as the spectral code with $\tau = 0.014$, for grids with $N = 2^{16}$. Thus both algorithms are comparable for typical problems.

### III. FRONT PROPAGATION SPEED

In this Section we will consider only the propagation of disturbances into a chaotic one dimensional system. As we said, essentially the same problems arise for front propagation into unstable states. Also, the extention to multi-dimensional system with steady or chaotic states would be straightforward.

In order to measure the propagation of a disturbance, we consider two realizations of the same field, say $A(x,t)$ and $B(x,t)$, which differ only in a finite region of space near the origin. If these states are unstable, the difference $\Delta A(x,t) = |A(x,t) - B(x,t)|$ will spread with a limiting velocity $v_F$. This velocity can be measured by recording the distance $R(t)$ of the *front*. This is the furthest point where $\Delta A(x,t)$ is not exactly equal to zero,

$$v_F = \lim_{t \to \infty} \frac{R(t)}{t}. \qquad (14)$$

This definition seems difficult to implement in practice, because we can only study finite systems. If we use periodic boundary conditions, the two opposing fronts (with velocities $v_F$ and $-v_F$) will collide with each other after $\approx L/2v_F$ time units. With open boundary conditions, the situation is not much better. In order to avoid this problem, we proceeded as follows. Assume that we have discretized space and time, and we know that the front cannot proceed by more than $m$ sites per time step (in our method, $m = (2N_c - 1)/2$). Then we take configuration $A$ as reference configuration, and replace in each step $B(x,t)$ by

$$B(x,t) \leftarrow A(x,t) \quad \forall x \in \, ]R(t), R(t) + m] \quad . \qquad (15)$$

By "cleaning" the solution $B$ in the region ahead of the front, we make sure that the front can propagate unperturbed, while the effect of this cleaning on the trailing end of the front should be negligible for sufficiently large systems.

Let us now assume that we want to use a spectral method. Due to the non-local nature of the Fourier transform, the difference $\Delta A(x,t)$ will not remain *identically* zero ahead of the front. This means that the perturbation will spread instantaneously, albeit with very small amplitude (typically of the order of the machine precision). Since the states are unstable, this tiny error will increase exponentially, preventing thereby any accurate measurement of $v_F$.

Speeds measured with our algorithm are given in table I, for $c_1 = 3.5$ and several values of $c_3$ (third column). Together with the speed, we measured also the average profile of the front of $\Delta A(x,t)$ in a co-moving coordinate system. Of course, since both $A$ and $B$ are chaotic, also $\Delta A$ will fluctuate. But we expect [5,6] that $\log \Delta A(x,t)$ will decay in average exponentially in space,

$$\Delta A(x,t) \sim e^{-\mu_F x} \quad . \tag{16}$$

Estimates for the exponent $\mu_F$ are given in the second column of table I.

Recently, it has beeen shown for coupled map lattices [15] that $\mu_F$ and $v_F$ are related [6]. More precisely, let us define the *specific Lyapunov exponent* $\lambda(\mu)$ [16] as the average growth in time, at fixed position $x$, of an infinitesimal perturbation $\delta A(x,t)$ which satisfies $\delta A(x,t) \sim e^{-\mu x}$ throughout the entire system (this can be achieved by special boundary conditions, $\delta A(L,t) = e^{-\mu L}\delta A(0,t)$). Furthermore, we define

$$V(\mu) = \frac{\lambda(\mu)}{\mu} \quad . \tag{17}$$

Then we have [6]

$$v_F = V(\mu_F) . \tag{18}$$

In addition, we know that $V(\mu)$ has a unique minimum for chaotic systems, so that [6]

$$v_F \geq v_L \equiv \min_\mu V(\mu) . \tag{19}$$

The speed $v_L$ is called the *linear velocity*, since it is obtained from a linear stability analysis.

The values of $v_L$ and of $\mu_L$ (the value of $\mu$ where $V(\mu)$ has its minimum) are given in the last two columns of table I. We see that they agree within errors with $v_F$ and $\mu_F$, respectively. Some discrepancies are observed between the spatial profiles $\mu_L$ and $\mu_F$ (in particular, for $c_3 = 0.9$), due to the difficulties to perform a direct measurement of the shape of the propagating front.

We see thus that disturbances are "pulled" [17] for the present parameter values of the CGLE, except maybe for the smallest values of $c_3$. That is, it is the very edge of the perturbation front (where $\Delta A$ is infinitesimal and described by linear stability analysis) which "pulls" it. In contrast, there are cases possible where the front is "pushed" by regions where the perturbation is finite. We will present in a forthcoming paper [18] evidence for pushed error propagation in the CGLE at different parameter values.

## IV. CONCLUSIONS

In the present paper we have introduced a new technique to integrate PDE's. It consists of a time-splitting procedure (as, e.g., the well known leap frog) where the integration of the operator which is non-local in configuration space is performed by convolution with its kernel.

This algorithm is particularly useful for propagation phenomena into unstable (and chaotic, in particular) states which cannot be treated by pseudospectral methods. We applied it to the propagation of disturbances in the CLGE, but we suggest that it will be useful also in many similar phenomena.

The speed of disturbance propagation is a relevant indicator to characterize spatially extended systems, being able not only to account for local chaoticity (associated to a maximal positive Lyapunov exponent) but also for the information spreading in the system. In the present paper, the situation has been considered where the propagation is essentially ruled by pulling (i.e., by a linear mechanism). This seems not to be always the case. The possibility of having different propagation mechanisms should be closely related to various phases observed for the CLGE (in particular, to the two different chaotic regimes found for this system, namely phase and defect turbulence) [7], and should help to explain the dynamics in this model. Work along these lines is in progress and will be reported elsewhere [18].

## ACKNOWLEDGMENTS

We want to thank A. Politi for useful suggestions and discussions, as well as D.A. Egolf and H.S. Greenside for providing us some unpublished data concerning the CGLE. One of us (A.T.) gratefully acknowledges the European Economic Community for the fellowship no ERBCHBICT941569 "Multifractal Analysis of Spatio-Temporal Chaos". This work was partly supported by DFG within the Graduiertenkolleg "Feldtheoretische und numerische Methoden in der Elementarteilchen- und Statistischen Physik". Finally, we want to thank the Institute for Scientific Interchange (I.S.I) in Torino (Italy) for the hospitality offered us during July 1995 within a workshop organized by the EEC-Network "Complexity and Chaos".

FIGURES

**Fig.1**: Real (solid line) and complex (dashed line) parts of the convolution kernel $\phi(x)$ for the CGLE (see eq.(13)), for $c_1 = 3.5$, $\Delta x = 0.098$ and $\tau = 0.05$.

**Fig.2**: Integration errors as a function of $N_c$ for several integration time steps $\tau$. The errors are estimated by comparing with a standard pseudo-spectral code with a smaller time step $\tau' = \tau/100$. The values for $c_1$ and $\Delta x$ are the same as in fig.1, the other parameters are $c_3 = 0.9$ and $N = 2048$. The errors shown are averages over several runs of duration $T = 0.2$ each.

TABLES

TABLE I. Propagation speed and slope of the exponential profile of the propagating front for the CGLE with parameter $c_1 = 3.5$ at some values of $c_3$. $\mu_F$ and $v_F$ refer to a direct measurement through the algorithm here introduced. $\mu_L$ and $v_L$ refer to the minimum of the function $V(\mu)$ defined in Eq. (17). The direct measurement of the front propagation has been performed adopting the method here introduced with $N_c = 110$, $N = 2048$, $\tau = 0.05$, and $\Delta x = 0.098$. After a transient of 100,000 time steps, the front has been followed for a number of time steps ranging from 1,840,000 to 2,100,000. The Lyapunov analysis necessary to estimate $\mu_L$ and $v_L$ has been done employing a pseudo-spectral code with $N = 2048$, $\tau = 0.01$ and $\Delta x = 0.098$. After a transient of 100,000 time steps the dynamics of the system has been integrated for a number of steps ranging from 470,000 to 1,290,000. In both cases periodic boundary conditions have been employed.

| $c_3$ | $\mu_F$ | $v_F$ | $\mu_L$ | $v_L$ |
|---|---|---|---|---|
| 0.9 | $\sim 0.16$ | $2.09 \pm 0.01$ | $0.134 \pm 0.006$ | $2.13 \pm 0.07$ |
| 0.8 | $\sim 0.12$ | $1.34 \pm 0.01$ | $0.110 \pm 0.005$ | $1.38 \pm 0.05$ |
| 0.7 | $\sim 0.10$ | $0.95 \pm 0.01$ | $0.085 \pm 0.005$ | $0.96 \pm 0.04$ |
| 0.6 | $\sim 0.08$ | $0.62 \pm 0.01$ | $0.075 \pm 0.006$ | $0.65 \pm 0.03$ |